\documentclass[epj]{svjour}
\usepackage{graphics}
\begin{document}
\title{Flavor asymmetry of the nucleon sea in an unquenched quark model}
\author{Roelof Bijker\inst{1} \and Elena Santopinto\inst{2}
}                     
%
%
\institute{ICN-UNAM, AP 70-543, 04510 Mexico DF, Mexico \and 
INFN, Sezione di Genova, via Dodecaneso 33, 16164 Genova, Italy}
\date{Received: date / Revised version: date}
%
\abstract{We discuss the flavor asymmetry of the nucleon sea in an unquenched 
quark model for baryons in which the effects of quark-antiquark pairs ($u \bar{u}$, 
$d \bar{d}$ and $s \bar{s}$) are taken into account in an explicit form. 
It is shown that the inclusion of $q \bar{q}$ pairs leads to an excess 
of $\bar{d}$ over $\bar{u}$ quarks in the proton. 
\PACS{
      {14.20.Dh}{Protons and neutrons}   \and
      {12.39.-x}{Phenomenological quark models} \and
      {11.30.Hv}{Flavor symmetries} \and
      {11.55.Hx}{Sum rules}
     } 
} 
\maketitle
\section{Introduction}

The flavor content of the nucleon 
sea provides an important test for models of nucleon structure. A flavor 
symmetric sea leads to the Gottfried sum rule $S_G=1/3$ \cite{gsr}, whereas 
any deviation from this value is an indication of the $\bar{d}/\bar{u}$ 
asymmetry of the nucleon sea. The first clear evidence of a violation of 
the Gottfried sum rule came from the New Muon Collaboration (NMC) \cite{nmc}, 
which was later confirmed by Drell-Yan experiments \cite{DY} and a 
measurement of semi-inclusive deep-inelastic scattering \cite{hermes}. 
All experiments show evidence that there are more $\bar{d}$ quarks in the 
proton than there are $\bar{u}$ quarks. The experimental studies and 
theoretical ideas on the flavor asymmetric sea are summarized in several 
review articles \cite{review}.

In the constituent quark model (CQM), the proton is described in terms 
of a $uud$ valence-quark configuration. Therefore, the violation of the 
Gottfried sum rule implies the existence of higher Fock components (such 
as $uud-q \bar{q}$ configurations) in the proton wave function. Additional 
indications for the importance of multiquark components are provided by 
parity-violating electron scattering experiments, which have shown evidence 
for a nonvanishing strange quark contribution, albeit small, to the charge 
and magnetization distributions of the proton \cite{Acha}, and by CQM 
studies of baryon spectroscopy \cite{cqm}.  

\begin{figure}
\centering
\resizebox{0.3\textwidth}{!}{\includegraphics{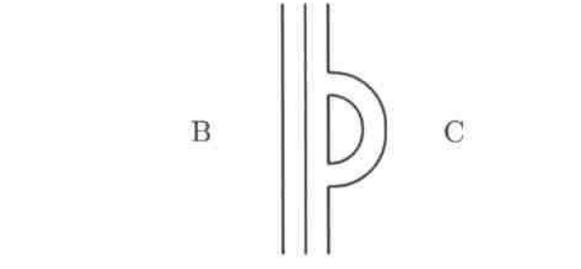}}
\caption{One-loop diagram at the quark level}
\label{fig:1}
\end{figure}

Theoretically, the role of $q^4 \bar{q}$ configurations in the nucleon 
wave function was studied in an application to the electromagnetic 
form factors \cite{riska}. Mesonic contributions to the spin and flavor 
structure of the nucleon are reviewed in \cite{review}. 
In another, CQM based, approach the importance of $s \bar{s}$ pairs in the 
proton was studied in a flux-tube breaking model based on valence-quark plus 
glue dominance to which $s \bar{s}$ pairs are added in perturbation \cite{baryons}. 
The pair-creation mechanism is inserted at the quark level and the one-loop 
diagrams are calculated by summing over a complete set of intermediate 
baryon-meson states $BC$ (see Fig.~\ref{fig:1}). 
The pairs are created with the $^{3}P_0$ quantum numbers of the vacuum. 
For consistency with the OZI-rule and to retain the success of the CQM in 
hadron spectroscopy, it was found necessary to sum over a complete set of 
intermediate states, including both pseudoscalar and vector mesons, rather 
than just a few low-lying states \cite{baryons,OZI}.  

In order to address the violation of the Gottfried sum rule, we have generalized 
the model of \cite{baryons} to include $u \bar{u}$ and $d \bar{d}$ loops as well. 
The formalism of the ensuing unquenched quark model is described in a separate 
contribution to these proceedings \cite{elena}. The aim of this manuscript 
is to discuss an application to the flavor asymmetry of the nucleon sea.  

\section{Flavor asymmetry}
 
The first clear evidence for the flavor asymmetry of the nucleon sea was provided 
by NMC at CERN \cite{nmc}. The flavor asymmetry is related to the Gottfried integral 
for the difference of the proton and neutron electromagnetic structure functions  
\begin{eqnarray*}
S_G &=& \int_0^1 dx \frac{F_2^p(x)-F_2^n(x)}{x} 
\nonumber\\ 
&=& \frac{1}{3} - \frac{2}{3} \int_0^1 dx \left[ \bar{d}(x) - \bar{u}(x) \right] ~.
\end{eqnarray*}
Under the assumption of a flavor symmetric sea $\bar{d}(x)=\bar{u}(x)$ one obtains the 
Gottfried sum rule $S_G=1/3$. The final NMC value is $0.2281 \pm 0.0065$ at $Q^2 = 4$ 
(GeV/c)$^2$ for the Gottfried integral over the range $0.004 \leq x \leq 0.8$ \cite{nmc}, 
which implies a flavor asymmetric sea. The violation of the Gottfried sum rule has been 
confirmed by other experimental collaborations \cite{DY,hermes}. Table~\ref{tab:1} 
shows that the experimental values of the Gottfried integral are consistent with each 
other within the quoted uncertainties, even though the experiments were performed 
at very different scales, as reflected in the average $Q^2$ values.  
Theoretically, it was shown that in the framework of the cloudy bag model  
the coupling of the proton to the pion cloud provides a mechanism to produce a flavor 
asymmetry due to the dominance of $n \pi^+$ among the virtual configurations \cite{Thomas}.  

In the unquenched quark model, the flavor asymmetry can be calculated from the 
difference of the number of $\bar{d}$ and $\bar{u}$ sea quarks in the proton 
\begin{eqnarray*}
N_{\bar{d}}-N_{\bar{u}} 
= \int_0^1 dx \left[ \bar{d}(x) - \bar{u}(x) \right] ~. 
\label{asym}
\end{eqnarray*}
Even in absence of explicit information on the (anti)quark distribution 
functions, the integrated value can be obtained directly from the left-hand side 
of Eq.~(\ref{asym}). The effect of the quark-antiquark pairs on the Gottfried 
integral is a reduction of about one third with respect to the Gottfried sum 
rule, corresponding to an excess of $\bar{d}$ over $\bar{u}$ quarks in the proton 
which is in qualitative agreement with the NMC result. 
It is important to note that in this calculation the parameters were taken from 
the literature \cite{baryons,CR}, and that no attempt was made to optimize their 
values. Due to isospin symmetry, the neutron has a similar excess of $\bar{u}$ 
over $\bar{d}$ quarks. 

\begin{table}
\caption{Experimental values of the Gottfried integral}
\label{tab:1}
\begin{tabular}{lccc}
\hline\noalign{\smallskip}
Experiment & $\langle Q^2 \rangle$ & $x$ range & $S_G$ \\
\noalign{\smallskip}\hline\noalign{\smallskip}
NMC & $4$ & $0.004 < x < 0.80$ & $0.2281 \pm 0.0065$ \\
HERMES & $2.3$ & $0.020 < x < 0.30$ & $0.23 \pm 0.02$ \\
E866/NuSea & $54$ & $0.015 < x < 0.35$ & $0.255 \pm 0.008$ \\
\noalign{\smallskip}\hline
\end{tabular}
\end{table}

\section{Summary, conclusions and outlook}

In this contribution, we discussed the importance of quark-antiquark pairs 
in baryon spectroscopy. The calculations were carried out in an unquenched 
quark model for baryons in which the contributions from $u \bar{u}$, $d \bar{d}$ 
and $s \bar{s}$ loops are taken into account in a systematic way \cite{elena}. 

The model was applied to the flavor asymmetry of the nucleon sea. In a first, 
exploratory, calculation in which the parameters were taken from the literature 
\cite{baryons,CR}, it was shown that the inclusion of $q \bar{q}$ pairs leads 
immediately to an excess of $\bar{d}$ over $\bar{u}$ quarks in the proton. We 
emphasize again that no attempt was made to optimize the parameters in the 
calculations. 

In our opinion the first results for the flavor asymmetry (discussed here) and 
the proton spin (see \cite{elena}) are very promising and encouraging. We believe 
that the inclusion of the effects of quark-antiquark pairs in a general and 
consistent way, as suggested in \cite{elena} and in this contribution, may provide 
a major improvement to the constituent quark model, increasing considerably its 
range and applicability. 

In future work, the unquenched quark model will be applied systematically to 
several problems in light baryon spectroscopy, such as the electromagnetic and 
strong couplings, the elastic and transition form factors of baryon resonances, 
their sea quark content and their flavor decomposition \cite{BS}.  

\section*{Acknowledgments}
This work was supported in part by a grant from CONACYT, Mexico 
and in part by INFN, Italy.

\end{document}